\documentclass[aps,prl,reprint,superscriptaddress]{revtex4-1}
\usepackage[utf8]{inputenc}
\usepackage{hyperref}
\usepackage{graphicx}
\usepackage{amsmath}
\usepackage[version=3]{mhchem} 

\usepackage{xcolor}
\begin{document}

\title{Confinement Effects on the Nuclear Spin Isomer Conversion of H\textsubscript{2}O}

\author{Pierre-Alexandre Turgeon}
\affiliation{Département de chimie, Université de Sherbrooke, Sherbrooke, J1K 2R1, CANADA }

\author{Jonathan Vermette}
\affiliation{Département de chimie, Université de Sherbrooke, Sherbrooke, J1K 2R1, CANADA }

\author{Gil Alexandrowicz}
\affiliation{Schulich Faculty of Chemistry, Technion - Israel Institute of Technology, Technion City, Haifa 32000, ISRAEL}

\author{Yoann Peperstraete}
\affiliation{LERMA, Observatoire de Paris, PSL Research University, CNRS, Sorbonne Universités, UPMC Univ. Paris 06, F-75252 Paris, FRANCE}

\author{Laurent Philippe}
\affiliation{LERMA, Observatoire de Paris, PSL Research University, CNRS, Sorbonne Universités, UPMC Univ. Paris 06, F-75252 Paris, FRANCE}

\author{Mathieu Bertin}
\affiliation{LERMA, Observatoire de Paris, PSL Research University, CNRS, Sorbonne Universités, UPMC Univ. Paris 06, F-75252 Paris, FRANCE}

\author{Jean-Hugues Fillion}
\affiliation{LERMA, Observatoire de Paris, PSL Research University, CNRS, Sorbonne Universités, UPMC Univ. Paris 06, F-75252 Paris, FRANCE}

\author{Xavier Michaut}
\email{Xavier Michaut, Tel.:33 (0)1 44 27 44 74, Email:xavier.michaut@upmc.fr}
\affiliation{LERMA, Observatoire de Paris, PSL Research University, CNRS, Sorbonne Universités, UPMC Univ. Paris 06, F-75252 Paris, FRANCE}

\author{Patrick Ayotte}
\email{Patrick Ayotte,Tel.:819-821-7889, Email: patrick.ayotte@usherbrooke.ca}
\affiliation{Département de chimie, Université de Sherbrooke, Sherbrooke, J1K 2R1, CANADA }

%

\begin{abstract}
The mechanism for interconversion between the nuclear spin isomers (NSI) of H\textsubscript{2}O remains shrouded in uncertainties. The temperature dependence displayed by NSI interconversion rates for H\textsubscript{2}O isolated in an Argon matrix provides evidence that confinement effects are responsible for the dramatic increase in their kinetics with respect to the gas phase, providing new pathways for o-H\textsubscript{2}O$\leftrightarrow$p-H\textsubscript{2}O conversion in endohedral compounds. This reveals intramolecular aspects of the interconversion mechanism which may improve methodologies for the separation and storage of NSI en route to application ranging from magnetic resonance spectroscopy and imaging to interpretations of spin temperatures in the interstellar medium.
\end{abstract}

\maketitle

\section{Introduction}
Collections of molecules that display indiscernible spin-bearing nuclei behave as a mixtures of nuclear spin isomers (NSI). While at high temperatures, their relative abundance is dictated by nuclear spin statistics, the occurrence of the different NSI at low temperatures is controlled by a complex interplay of symmetry considerations, quantum states degeneracies and energetics in addition to their interconversion kinetics (i.e., sample thermal history). In hydrogen gas, NSI interconversion is sufficiently slow to allow their separation and storage allowing subsequent manipulations. Consequently, several methodologies are nowadays available for the preparation of molecular hydrogen samples highly enriched in either of its NSI \cite{farkas1935orthohydrogen, Fukutani2013},  thereby enabling powerful magnetic resonance applications \cite{Bowers1986}. However, the case of polyatomic molecules still presents significant challenges. While only very slight enrichments have thus so far been reported for simple molecules (i.e., CH\textsubscript{3}F, C\textsubscript{2}H\textsubscript{4}, H\textsubscript{2}CO) using laser-based methods \cite{Sun2005,Nagels1996,Peters1999}  much larger enrichments were reported for H\textsubscript{2}O using molecular beam \cite{Kravchuk2011,Turgeon2012a,Horke2014} or chromatographic methodologies \cite{Tikhonov2002}. Nevertheless, the paucity of general and efficient enrichment methodologies for the preparation of out-of-equilibrium spin populations in polyatomic molecules continues to hamper studies of their NSI interconversion mechanism and rates. Conversely, improving our understanding of NSI interconversion is required to devise efficient separation and storage strategies hence, significant gaps still remain in our description of these important fundamental, and intimately intertwined processes.

\begin{figure}
\includegraphics[width=1\columnwidth]{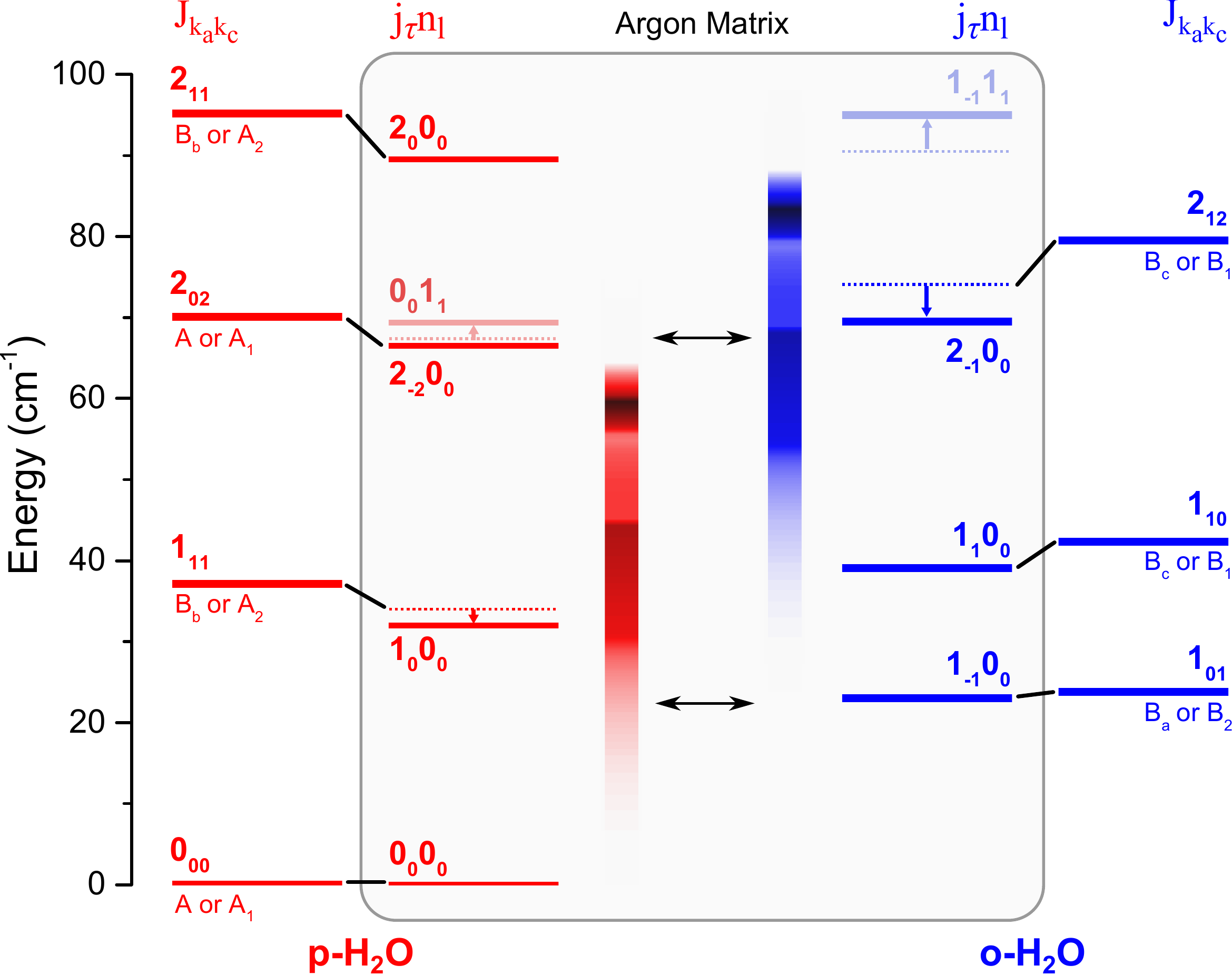}%
\caption{\label{figure1}Energy levels diagram of gas phase H\textsubscript{2}O \cite{Tennyson2013} (far left and right) and H\textsubscript{2}O@Ar \cite{Perchard2001a} (shaded region). The gas phase states are labeled using the asymmetric rotor convention $J_{k_a k_c}$ while the confined states are identified using the quantum numbers of an asymmetric rotor $j$ and $\tau$, where $\tau=k_a-k_c$, and those of a 3D isotropic harmonic oscillator $n$ and $l$. The double-headed arrows depict the NSI interconversion pathways highlighted in the present work. The phonon density of state of the argon crystal is displayed using vertical shaded bars alongside the o-\ce{H2O}@Ar and p-\ce{H2O}@Ar confined rotor energy levels (ref. \cite{Fujii1974,Kaburaki2007}).}
\end{figure}

In isolated water molecules, the total nuclear spin is ever so weakly coupled to the other molecular degrees of freedom resulting in that, for most practical purposes,  interconversion between ortho-H\textsubscript{2}O (i.e., o-H\textsubscript{2}O, whose rotational energy level are labeled using the conventional approximate quantum numbers J\textsubscript{k\textsubscript{a}k\textsubscript{c}}, and displayed to the far right of Figure \ref{figure1}) and para-H\textsubscript{2}O (i.e., p-H\textsubscript{2}O, Figure \ref{figure1}; far left) can often be considered as a forbidden process. Even in the absence of a spin conversion catalyst \cite{Wigner}, NSI interconversion may nonetheless arise in water vapour, albeit at a very slow rate, due to intermixing of very nearly degenerate o-H\textsubscript{2}O and p-H\textsubscript{2}O states\cite{Curl1967,Chapovsky1999} that couple through weak hyperfine interactions \cite{Flygare1974}. Incidentally, the coherence that may thus build between these weakly coupled states is eventually destroyed by external perturbations (i.e., intermolecular collisions), thereby causing the wavefunction of the mixed states to collapse and NSI interconversion to occur \cite{Curl1967,Chapovsky1999}. However, due to the magnitude of the intramolecular magnetic interactions (i.e., the spin-spin and spin-rotation coupling strengths are only on the order of a few tens of kHz) \cite{Cazzoli2009,Puzzarini2009c}, this so-called “quantum relaxation” mechanism should only be operative at elevated temperatures (whereby nearly degenerate, highly rotationally excited states acquire significant population) \cite{Curl1967,Chapovsky1999} and pressures (thereby increasing the rate of intermolecular collisions) \cite{Cacciani2012}.

Furthermore, while it can easily be shown that the dipole-dipole interaction between the protons’ nuclear spins cannot couple o-H\textsubscript{2}O and p-H\textsubscript{2}O states, symmetry considerations impose strict selection rules upon which states can be coupled by the other intramolecular magnetic interactions within the H\textsubscript{2}O molecule \cite{Curl1967}. Indeed, the spin-rotation (SR) Hamiltonian (which transforms as B\textsubscript{1} in the C\textsubscript{2v} point group) \cite{Curl1967} can only couple rotational states of certain symmetries, namely A\textsubscript{1}$\leftrightarrow$B\textsubscript{1} and A\textsubscript{2}$\leftrightarrow$B\textsubscript{2}. Consequently, at very low temperatures, ever so small contributions to the NSI interconversion rate by ways of the SR coupling are only to be expected between the 1\textsubscript{01} o-H\textsubscript{2}O state and the 1\textsubscript{11} (as well as 2\textsubscript{11}) p-H\textsubscript{2}O state while, at slightly higher temperatures, conversion may be mediated by SR couplings between the 2\textsubscript{02} p-H\textsubscript{2}O state and the 1\textsubscript{10} (as well as 2\textsubscript{12}) o-H\textsubscript{2}O state.  Interestingly, symmetry considerations forbid the coupling of any o-H\textsubscript{2}O state with the 0\textsubscript{00} ground state of p-H\textsubscript{2}O by ways of the SR interaction. Therefore, given our current grasp of the mechanism for NSI interconversion, it is thus widely believed that, under the cold and dilute conditions of the interstellar medium, the NSI lifetimes for isolated H\textsubscript{2}O molecules could reach billions of years \cite{Tanaka2013}.

NSI interconversion in H\textsubscript{2}O was reported to proceed at a much greater rate in the condensed phase (i.e., inert matrices, liquid water, amorphous or crystalline ice)\cite{Redington1963a,Pardanaud2007, Abouaf-Marguin2009, fillion_understanding_2012},  possibly mediated by inter-molecular spin-spin (SS) couplings \cite{Limbach2006,Buntkowsky2008}. However, there has also been some reports of highly stable out-of-equilibrium spin state populations in liquid water and in crystalline ice \cite{Tikhonov2002}. While at odds with the rapid NSI interconversion promoted by the inter-molecular SS coupling, these claims supported an intriguing proposition from the astrophysical community, namely that the unexpectedly low spin “temperatures” in comas \cite{Mumma1986c,Crovisier1997} and protoplanetary disks \cite{Hogerheijde2011} could be vestiges of the conditions of formation of these celestial icy bodies and/or of their constitutive molecular material, thereby providing a valuable “astronomical clock” \cite{Hama2011}. This hypothesis raises the interesting question of how the populations of H\textsubscript{2}O spin states in ice are related to those of the adsorbing/desorbing water molecules \cite{Hama2011,Sliter2011,Hama2013,Hama2016}. Clearly, a better understanding of the (intra- and inter-molecular) NSI interconversion mechanisms and rates is required in order to improve the interpretation of astrophysical proxies while it may also contribute to develop better methodologies for the separation\cite{Kravchuk2011,Horke2014}  and storage\cite{Turgeon2012a} of o-H\textsubscript{2}O, opening up prospects for orders of magnitude enhancements in sensitivity for magnetic resonance spectroscopy and imaging\cite{Bowers1986,Kravchuk2011}.

\section{Experimental Section}
Confinement effects on the NSI interconversion mechanism and rates for matrix isolated H\textsubscript{2}O@Ar were examined by scrutinizing the o-H\textsubscript{2}O$\leftrightarrow$p-H\textsubscript{2}O interconversion kinetics at temperatures between 4K and 25K using infrared spectroscopy. Matrix isolated H\textsubscript{2}O, a well-known \cite{Redington1963a} class of endohedral H\textsubscript{2}O compounds along with H\textsubscript{2}O@\ce{C60} \cite{Kurotobi2011,Beduz2012}, was prepared by slowly condensing a mixture of water vapor and argon gas onto a gold-plated copper substrate at 20K. These conditions are known to result in \ce{H2O} molecules being confined to substitutional sites in a crystalline rare gas matrix \cite{Langel1988b,Knoezinger1983,Knozinger1988a}. 
Under our experimental conditions, argon has been shown to adopt a Face-Centered Cubic (FCC, which displays O\textsubscript{h} symmetry substitutional site) crystal structure while the occurrence of the metastable Hexagonal Close-Pack (HCP, which displays D\textsubscript{3h} symmetry substitutional site) crystal structure may be promoted by impurities \cite{Meyer1964}. In contrast to \ce{CH4}@Ar,\cite{jones_hindered_1986} where distinctive spectral features (i.e., the crystal field splitting of spectral features due to the degeneracy of the rovibrational sub-levels being lifted in the D\textsubscript{3h} symmetry HCP sites) and different NSI interconversion kinetics could be attributed to molecules trapped in HCP and FCC sites, neither could be observed in \ce{H2O}@Ar.\cite{Redington1963a, Pardanaud2007, Abouaf-Marguin2009}. At a dilution ratio of 1 H\textsubscript{2}O:1000 Ar, contributions from inter-molecular couplings \cite{Limbach2006,Buntkowsky2008} to the NSI interconversion rates of H\textsubscript{2}O have been shown to be negligible\cite{Pardanaud2007,Abouaf-Marguin2009} allowing intra-molecular contributions to be highlighted.  

\section{Results and Discussion}
The shaded region in Figure 1 displays the rotational-translational (RT) energy levels for H\textsubscript{2}O@Ar,\cite{Perchard2001a} which were also calculated by Ceponkus et al \cite{Ceponkus2013e}, according to the toy model originally proposed by Friedman and Kimmel \cite{Friedmann1967b}. While it reveals that H\textsubscript{2}O molecules rotate relatively “unhindered”, confinement of H\textsubscript{2}O to a substitutional site of a rare gas matrix has three important consequences. Firstly, rotational motion of the \ce{H2O} molecule does not proceed around its center-of-mass (COM), but rather around its center-of-interaction (COI) with the Ar matrix.  The energy level diagram for the confined quantum asymmetric rotor can thus be most simply interpreted using effective rotational constants A and C for H\textsubscript{2}O@Ar that are significantly smaller than those of free H\textsubscript{2}O (effective confined rotor states are linked to their corresponding free rotor states in Figure \ref{figure1}).  This simple model allows one to estimate that the COI is located (0,15$\pm$0.01 \AA) away from the COM of H\textsubscript{2}O and that it lies along its C\textsubscript{2}-axis. Secondly, H\textsubscript{2}O@Ar displays local oscillator (LO) states (reported in Figure \ref{figure1} as lighter shaded levels near 68 cm\textsuperscript{-1} for p-H\textsubscript{2}O@Ar and near 95 cm\textsuperscript{-1} for o-H\textsubscript{2}O@Ar) resulting from quantization of the frustrated translational motions of H\textsubscript{2}O confined to interstitial sites in the Ar crystal\cite{Pitsevich2015}. Finally, rotation around the COI results in a strong coupling between rotational and (frustrated) translational motions of confined H\textsubscript{2}O@Ar.  Despite evidence \cite{Knoezinger1983,Fry1984,Michaut2004} for significant RT coupling in H\textsubscript{2}O@Ar, confined rotor states are nonetheless labeled using the (uncoupled) asymmetric rotor quantum numbers $j$ and $\tau$, and the isotropic 3-D harmonic oscillator quantum numbers $n$ and $l$ (i.e., $j_\tau n_l$) for convenience. An important consequence of confinement, as it relates to NSI interconversion, is therefore, the strong mixing, and thus repulsion, between the 1\textsubscript{0}0\textsubscript{0} (rotational) and 0\textsubscript{0}1\textsubscript{1} (LO) states of p-H\textsubscript{2}O@Ar, and between the 2\textsubscript{-1}0\textsubscript{0} (rotational) and 1\textsubscript{-1}1\textsubscript{1} (LO) states of o-H\textsubscript{2}O@Ar, as a result of the RT coupling.

\begin{figure}
\includegraphics[width=1\columnwidth]{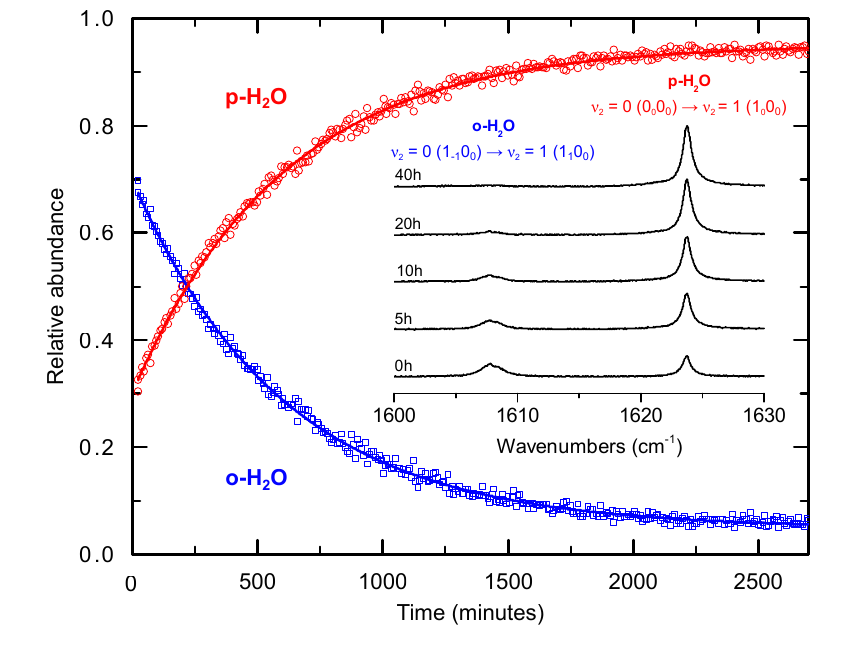}%
\caption{\label{figure2}NSI interconversion of \ce{H2O}@Ar at 6K causes  ortho-\ce{H2O} (open-squares) to convert to para-\ce{H2O} (open-circles) and the OPR to decay from $\sim$1.7 to \textless 0.1. The red and blue lines show the excellent agreement with a reversible first-order kinetics model.}
\end{figure}

Due to the quasi-free rotor character of H\textsubscript{2}O@Ar, it is straightforward to probe their nuclear spin state populations using ro-vibrational spectroscopy \cite{Pardanaud2007,Abouaf-Marguin2009,Redington1963a,fillion_understanding_2012}. This is illustrated in the inset to Figure \ref{figure2}, which reports the temporal evolution of the intramolecular HOH bending vibration (i.e., $\nu_2$ mode) infrared spectral range of H\textsubscript{2}O@Ar following a sudden decrease in sample temperature (T = 20K $\rightarrow$ 6K, at t = 0).  The intensity of the transitions assigned to o-H\textsubscript{2}O decreases, while that assigned to p-H\textsubscript{2}O increases, as a result of NSI interconversion throughout the 45h duration the NSI interconversion kinetics were monitored at T = 6K. The relative oscillator strengths for the ro-vibrational transitions arising from both NSI were evaluated \cite{Pardanaud2007,Abouaf-Marguin2009,Michaut2004} allowing the time evolution of their relative abundances to be quantified as reported in Figure \ref{figure2} for T = 6K (o-H\textsubscript{2}O - open squares ; p-H\textsubscript{2}O - open circles).

\begin{figure}
\begin{minipage}{\linewidth}
\includegraphics[width=1\columnwidth]{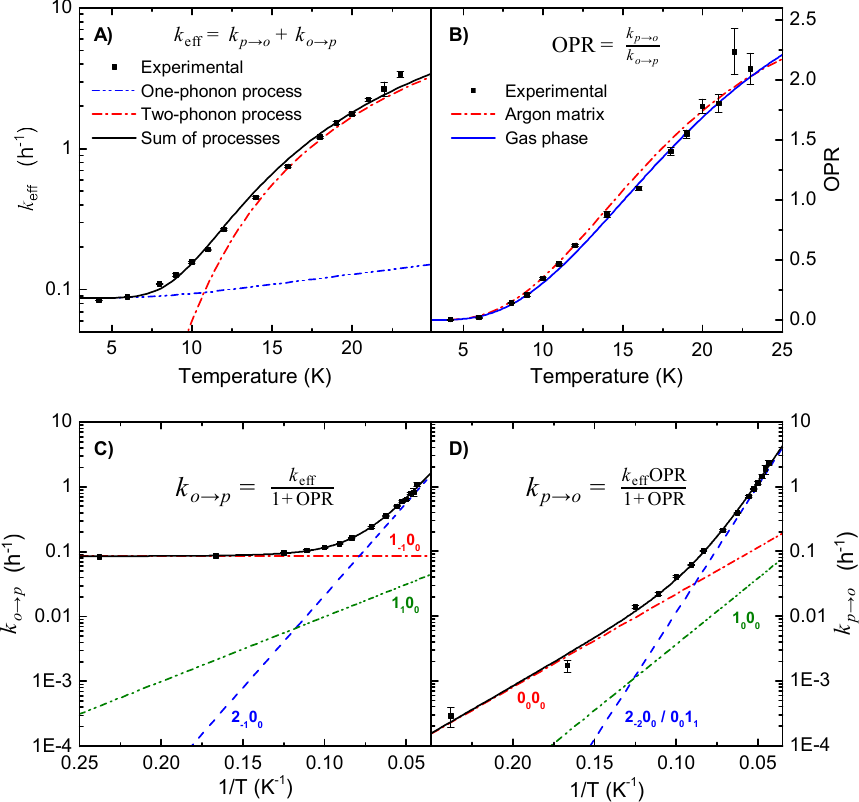}%
\caption{\label{figure3}(A) The experimental effective first-order rate constants (black squares) are compared to the empirical spin-lattice models proposed in ref. \cite{Scott1962}. (B) The experimentally measured OPRs (black squares), which agree with the gas phase OPR\protect\footnote{OPR for H\textsubscript{2}O were calculated using the most recent IUPAC data \cite{Tennyson2013}.} and the calculated OPR for \ce{H2O}@Ar\protect\footnote{OPR was calculated for H\textsubscript{2}O@Ar using the energy levels from ref. \cite{Perchard2001a}, displayed in Figure \ref{figure1}, which summarizes results from FIR and MIR spectroscopies, as well as INS (Ref \cite{Pardanaud2007, Abouaf-Marguin2009, Langel1988b,Knoezinger1983,Knozinger1988a,Ceponkus2013e}).}, allowed the calculation of the reversible rate constants $k_{o\rightarrow p}$ (C) and $k_{p\rightarrow o}$ (D) from $k_{\text{eff}}$ (A) assuming interconversion proceeds by reversible first-order kinetics.  Individual contributions from specific RT levels to the interconverison rate could thus be deduced exclusively from experimental data.}
\end{minipage}
\end{figure}

Assuming o-H\textsubscript{2}O$\leftrightarrow$p-H\textsubscript{2}O interconversion proceeds through reversible first-order kinetics (Figure \ref{figure2}, continuous lines), the effective interconversion rate constant ($k_{\text{eff}}$; Figure \ref{figure3}A), as well as the equilibrium constant (expressed as the asymptotic ortho-to-para ratio as t$\rightarrow\infty$, or OPR; Figure \ref{figure3}B), were obtained at several temperatures from 4K and up to 25K. Figure \ref{figure3}A shows that $k_{\text{eff}}$ displays a weak temperature dependence up to $\sim$10K, but that it increases rapidly thereafter (until reliable rate constants could no longer be obtained at T\textgreater 25K due to poor signal-to-noise ratio and increasingly rapid NSI interconversion).  

Phonons of the Ar matrix, whose density of states is indicated as vertical shaded bars in Figure \ref{figure1} \cite{Fujii1974,Kaburaki2007}, play an important role in the NSI interconversion mechanism which can be described rather well by empirical spin-lattice relaxation models (Figure \ref{figure3}A, continuous line) \cite{Scott1962,Miyamoto2008,fillion_understanding_2012,ueta_surface_2016}. At T\textless10K, the contribution from a reversible one-phonon process [i.e., the so-called “direct process” o-H\textsubscript{2}O$\leftrightarrow$p-\ce{H2O}+$\hbar\omega_{\text{Ar}}$, whose temperature behavior scales as $\text{coth}(\hbar\omega_{\text{Ar}}/2RT)$ with $\hbar\omega_{\text{Ar}}\sim 23$ cm\textsuperscript{-1}] appears to dominate NSI interconversion rates (Figure \ref{figure3}A, blue dash-dot-dot line). At T\textgreater10K, contributions from reversible two-phonons processes, o-H\textsubscript{2}O+$\hbar\omega'_{\text{Ar}}\leftrightarrow$ p-\ce{H2O}+$\hbar\omega''_{\text{Ar}}$, to the NSI interconversion rates become increasingly important.  While the temperature dependence of the experimental data is coherent with a resonant two-phonons Orbach process [i.e., which scales as $\text{exp}(-\Delta/RT)$ with $\Delta\sim46$cm\textsuperscript{-1}] (Figure \ref{figure3}A, red dash-dot line), a contribution from the non-resonant two-phonons Raman process [which scales as $T^n$, with $n\sim7$, and showed to be dominant in case of \ce{H2} adsorbed onto Amorphous Solid Water \cite{ueta_surface_2016}]  cannot be ruled out [and is even expected at higher temperatures which would, hence, better describe the temperature dependence of the conversion rate for T\textgreater20K].  Finally, the temperature dependence displayed by the spectroscopically determined OPR (Figure \ref{figure3}B - full square) agrees quite well with that calculated, using the energy level diagrams of Figure \ref{figure1},  for gas-phase H\textsubscript{2}O (continuous line)\cite{Tennyson2013}  and for confined H\textsubscript{2}O@Ar (dashed line) \cite{Ceponkus2013e,Perchard2001a}.

Pressing on within the assumption that NSI interconversion proceeds through reversible first-order kinetics and exploiting detailed balance, the first-order rate constants for the o-H\textsubscript{2}O$\rightarrow$p-H\textsubscript{2}O ($k_{o\rightarrow p}$) and p-H\textsubscript{2}O$\rightarrow$o-H\textsubscript{2}O ($k_{p\rightarrow o}$) unimolecular first-order half-reactions were calculated from the experimentally determined $k_{\text{eff}}$ (Figure \ref{figure3}A) and OPR (Figure \ref{figure3}B). The Arrhenius plots for $k_{o\rightarrow p}$ (Figure \ref{figure3}C) and $k_{p\rightarrow o}$ (Figure \ref{figure3}D) reveal clearly that the NSI interconversion kinetics in H\textsubscript{2}O@Ar display two distinct temperature regimes. For temperatures below about 10K (i.e., $T^{-1}>0.1$), the o-H\textsubscript{2}O$\rightarrow$p-H\textsubscript{2}O half-reaction (Figure \ref{figure3}C) appears barrier-less ($E_{a,\text{low }T}^{o\rightarrow p} = 0.6\pm0.3$cm\textsuperscript{-1}), whereas the p-H\textsubscript{2}O$\rightarrow$o-H\textsubscript{2}O half-reaction (Figure \ref{figure3}D) displays an apparent activation energy, $E_{a,\text{low }T}^{p\rightarrow o}= 22\pm$2 cm\textsuperscript{-1}. Above about 10K (i.e., $T^{-1}<0.1$), the o-H\textsubscript{2}O$\rightarrow$p-H\textsubscript{2}O (Figure \ref{figure3}C) and p-H\textsubscript{2}O$\rightarrow$o-H\textsubscript{2}O (Figure \ref{figure3}D) half-reactions display apparent activation energies $E_{a,\text{high }T}^{o\rightarrow p}= 46\pm2$ cm\textsuperscript{-1} and $E_{a,\text{high }T}^{p\rightarrow o} = 58\pm4$ cm\textsuperscript{-1}, respectively. These apparent activation energies are displayed as double-headed arrows on the energy diagram for H\textsubscript{2}O@Ar in Figure \ref{figure1}.

In the spirit of the “quantum relaxation” model \cite{Curl1967, Chapovsky1999},this may help pinpoint the gateway RT states that mediate NSI interconversion in H\textsubscript{2}O@Ar thereby providing invaluable insight into the underlying mechanism. Accordingly, the empirical rate constants $k_{o\rightarrow p}$ (Figure \ref{figure3}C) and $k_{p\rightarrow o}$ (Figure \ref{figure3}D) were described by a sum of Boltzmann-weighted contributions from individual RT states (Figure \ref{figure1})\cite{Curl1967}.  This allows to determine that, at T\textgreater10K, interconversion proceeds mostly through coupling between the $2_{-1}0_0$ RT state of o-H\textsubscript{2}O and the $2_{-2}0_0$/$0_01_1$ RT states of p-H\textsubscript{2}O and that their contributions to the o-\ce{H2O}$\rightarrow$p-\ce{H2O} and p-\ce{H2O}$\rightarrow$o-\ce{H2O} rates are ($13.0\pm0.2$) h\textsuperscript{-1} and ($85\pm30$) h\textsuperscript{-1}, respectively. Here, confinement effects manifest themselves most prominently through the strong RT coupling between confined o-\ce{H2O} states which display the same total angular momentum $J=j+l=2$ \cite{Ceponkus2013e}, namely the $2_{-1}0_0$ (mostly rotational) and $1_{-1}1_1$ (mostly LO) states of o-H\textsubscript{2}O. As a result, the $2_{-1}0_0$ o-H\textsubscript{2}O state acquires LO character and is significantly stabilized whereby it becomes close enough in energy to the $0_01_1$ (LO) and $2_{-2}0_0$ (rotational) states of p-H\textsubscript{2}O@Ar to allow interactions through the weak spin-rotation coupling thereby enhancing NSI interconversion. This interpretation is currently being validated using isotope substitution (H\textsubscript{2}\textsuperscript{A}O, A=16,17,18) and different confining medium (Ne, Ar, Kr, Xe) allowing to systematically tune the rotational energy levels of H\textsubscript{2}O@Ar and to modulate the strength of the RT coupling. Nonetheless, additional contributions arising from naturally occurring magnetic isotopes (\textsuperscript{21}Ne, \textsuperscript{83}Kr, \textsuperscript{129}Xe, \textsuperscript{131}Xe, \textsuperscript{17}O) need to be accounted for since they open new intermolecular NSI interconversion pathways.

The most intriguing finding reported herein remains, however, the unexpectedly rapid NSI interconversion observed at T\textless10K. Indeed, while interconversion at these temperatures is expected to be extremely inefficient in free H\textsubscript{2}O due to the strongly forbidden character of the intramolecular SS and SR magnetic couplings between the $0_{00}$ p-H\textsubscript{2}O and $1_{01}$ o-H\textsubscript{2}O ground states\cite{Tanaka2013}, comparatively large rates are observed in H\textsubscript{2}O@Ar. Furthermore, the weak temperature dependence displayed by $k_{\text{eff}}$ at T\textless10K (Figure \ref{figure3}A) strongly suggests creation and annihilation of $\hbar \omega\sim23$ cm\textsuperscript{-1} phonons in the Ar matrix and their scattering with confined H\textsubscript{2}O can promote NSI interconversion at an appreciable rate. The low temperature behaviors of $k_{\text{eff}}$ and OPR allow the contributions to NSI interconversion from the direct mechanism (i.e., o-H\textsubscript{2}O$\leftrightarrow$p-H\textsubscript{2}O+$\hbar\omega_{\text{Ar}}$) to be estimated as ($0.085\pm0.003$) h\textsuperscript{-1} from the $1_{-1}0_0$ o-H\textsubscript{2}O ground state and ($0.6\pm0.1$) h\textsuperscript{-1} from the $0_00_0$ p-H\textsubscript{2}O ground state\footnote{Upper limits to the rate from the $1_10_0$ o-H\textsubscript{2}O state and from the $1_00_0$ p-H\textsubscript{2}O states are estimated to be ($0.13\pm0.06$) h\textsuperscript{-1} and ($0.7\pm0.9$) h\textsuperscript{-1} respectively, however, their contributions to the empirical rate constants is no more than 20\%.}. Here, confinement effects must allow to overcome the strongly forbidden character of the hyperfine SS and SR couplings to the p-\ce{H2O} ground state. This might arise from the mixing between RT states of the C\textsubscript{2v} asymmetric quantum rotor due to trapping in the O\textsubscript{h} crystal field of cuboctaedral geometry of an argon matrix substitutional site. Indeed, Momose and coworkers previously explored crystal field effect as they pertained to the NSI interconversion kinetics displayed by \ce{CD4}/\ce{CH4}@p-\ce{H2}\cite{Miyamoto2008}. They also invoked magnetic interactions of higher orders than spin-spin or spin-rotation couplings which remain highly forbidden for methane in p-\ce{H2} matrices.  The situation, however, is significantly more complex in \ce{H2O}@Ar due to couplings between angular momenta for intramolecular rotation, orbital motion of the 3D harmonic oscillator, and the precession of the water molecules' center of mass around the cavity. Therefore, in addition to phonon scattering, these intricate angular momentum coupling schemes arising from quantum confinement effects in \ce{H2O}@Ar must open novel NSI interconversion pathways that do not exist in free \ce{H2O}.

\begin{figure}
\includegraphics[width=1\columnwidth]{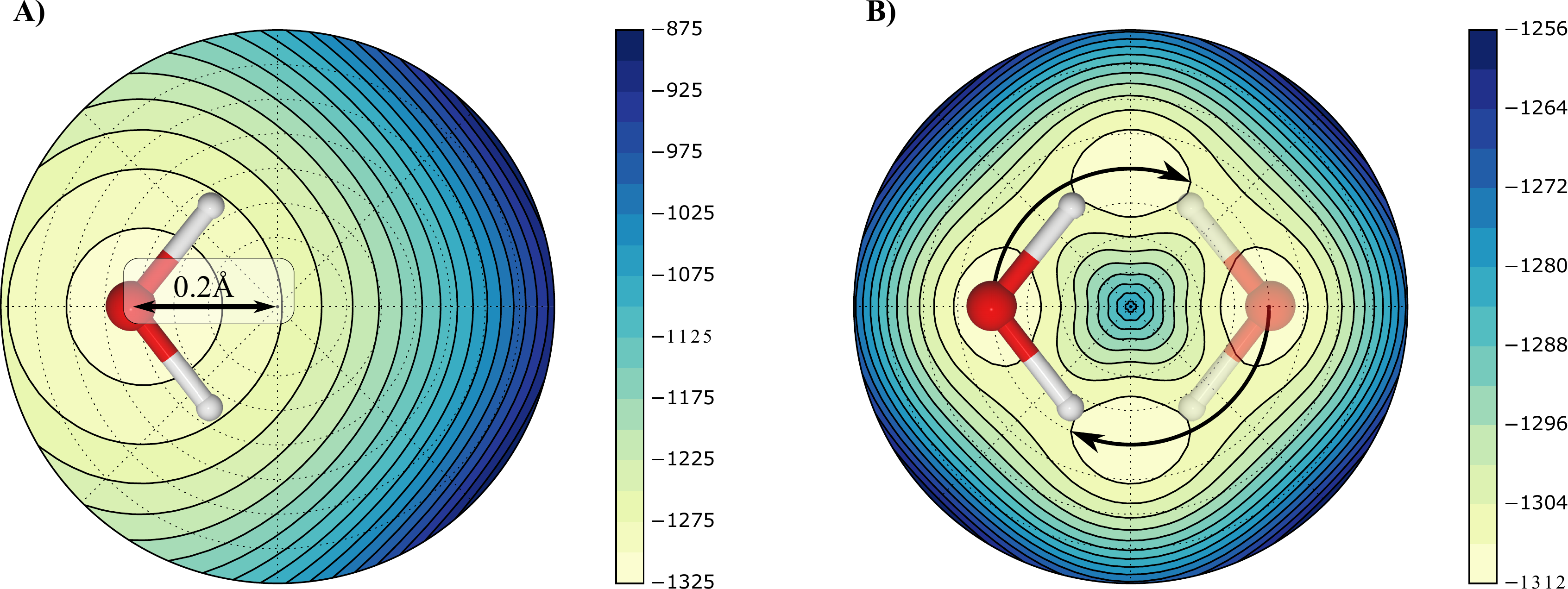}%
\caption{\label{figure4}Potential energy surface (cm\textsuperscript{-1}) resulting from the interaction of an H\textsubscript{2}O molecule with the 12 nearest neighbors of an Ar face-centered cubic (FCC) crystal substitutional site. The potential was calculated using a pairwise interaction potential\cite{Makarewicz2008a,Cohen1991a}. Calculations using an HCP structure reproduced the main features of the potential, while displaying only subtle differences. Panel A) presents the interaction potential for a water molecule moving along the [100] plane with its orientation fixed with respect to the face-centered cubic argon crystal, thereby highlighting the nearly harmonic character of the confining potential. Panel B) presents the rotationally  diabatic interaction potential for a water molecule moving in the same plane [100] while maintaining an orientation where the C\textsubscript{2} axis points toward the center of the cavity (as a result of strong RT coupling).}
\end{figure}

Proper account for these confinement effects on the complex coupled rotational-translational motions in \ce{H2O}@Ar would require solving the full dimensional Hamiltonian for the confined asymmetric rotor. This was achieved for CO@C\textsubscript{60} by Olthof et al. \cite{Olthof1996}, for H\textsubscript{2}O@C\textsubscript{60} by Felker and Ba{\v{c}}i\'c \cite{Felker2016} as well as for \ce{H2}@\ce{C60}, HD@C\textsubscript{60} and \ce{D2}@\ce{C60} by Xu et al.\cite{Xu2008} These studies described how the rotation of the “incarcerated” molecule results in a very strong coupling between the angular momenta of the rotating molecule and that arising from the rotation of its center of mass around the center of interaction within the cavity thus providing the basis for strong RT coupling.  In order to garner deeper insight into the effects of quantum confinement and its role in mediating new and alternate pathways for NSI interconversion in H\textsubscript{2}O@Ar, the interaction potential between H\textsubscript{2}O and the argon matrix was explored using accurate (high-level electronic structure-based\cite{Makarewicz2008a} or spectroscopic\cite{Cohen1991a}) Ar-H\textsubscript{2}O pair potentials. As displayed in Figure \ref{figure4}, the most salient feature of this (only very slightly anisotropic) potential for H\textsubscript{2}O@Ar is that the center-of-mass of the water molecule does not coincide with the geometric center of the cavity. As displayed in Figure \ref{figure4}A, the minimum energy configurations, in any given orientation of the water molecule, are found when the COM of H\textsubscript{2}O recedes $\rho\approx0.2$\AA{} away from the geometric center of the cavity along its C\textsubscript{2} (i.e, rotational b) axis. Furthermore, the frequency of the (uncoupled 3-D harmonic) LO mode for H\textsubscript{2}O@Ar was estimated to be $\omega_{\text{LO}}\approx60$ cm\textsuperscript{-1}. These values are in reasonable agreement with those obtained by Ceponkus et al. (i.e., $\rho\approx0.1$\AA, and $\omega_{\text{LO}}\approx68$ cm\textsuperscript{-1})\cite{Ceponkus2013e} considering the simplicity of their model.  The topology of the confining potential therefore possess characteristics that should result in much stronger mixing between the rotational and (frustrated) translational motions in H\textsubscript{2}O@Ar than that reported, for example, for H\textsubscript{2}O@C\textsubscript{60}.\cite{Olthof1996} Furthermore, while inter-molecular coupling between \ce{H2O} and the stiff fullerene cage were neglected in H\textsubscript{2}O@C\textsubscript{60}, the role of the soft phonon modes in H\textsubscript{2}O@Ar, which is responsible for the rapid thermalization of rotational motion and the efficient NSI interconversion, should be explicitly taken into account for a proper description of NSI interconversion in \ce{H2O}@Ar.

The temperature dependence of the effective first-order rate constant, $k_{\text{eff}}$, and equilibrium constant, OPR, enabled detailed analysis of which rotational-translation states contribute most to the interconversion mechanisms in \ce{H2O}@Ar.  Confinement effects provide dramatic enhancements in the o-\ce{H2O}$\leftrightarrow$p-\ce{H2O} interconversion rates in argon matrices.  Improving our understanding of confinement effect on the NSI interconversion mechanisms and rates should enable better storage strategies to be devised for water samples highly enriched in o-\ce{H2O} en route towards NMR applications to quantum information, spectroscopy and imaging. It may also provide more robust interpretations of the nuclear spin isomer populations observed in the interstellar medium. 

\section{Conclusion}
NSI interconversion kinetics of \ce{H2O} isolated in argon matrices was scrutinized using rovibrationnal spectroscopy at low temperature. The strong confinement effects observed spectroscopically in \ce{H2O}@Ar were shown to partially lift the rigorous selection rules that govern spin conversion in individual molecules, specially the forbidden coupling between the ortho and para ground states. Given the unprecedented quality of the data presented herein, this work brings new light to the long-studied problem of ortho/para conversion in \ce{H2O} by providing quantitative contributions of individual rotational-translational levels to the effective conversion rate. As this works is the first to provide state specific kinetics at low temperature, it constitutes a milestone in the elucidation of underlying NSI interconversion mechanisms.

\begin{acknowledgements}
Financial support by NSERC, CFI and CQMF is greatfully acknowledged. This work was supported by the French program funded by CNRS and CNES named “Physique et Chimie du Milieu Interstellaire” (PCMI) and the French National Research Agency (Project ANR GASOSPIN number 09-BLAN-0066-01). GA gratefully acknowledges funding from the German-Israeli Foundation for Scientific Research and the European Research Council under the European Unions seventh framework program (FP/2007-2013)/ERC grant 307267. PA gratefully acknowledges support by UPMC and CPCFQ during his sabbatical leave. Authors thank Karen Monneret and Julien Camperi for assistance. 
\end{acknowledgements}

\bibliography{bibliography}

\end{document}